\documentclass[runningheads]{llncs}
\usepackage[T1]{fontenc}
\usepackage{graphicx}
\usepackage{amsmath} 
\usepackage{graphicx}
\usepackage{subcaption}
\usepackage{booktabs}
\usepackage[many]{tcolorbox}
\usepackage{multirow}
\begin{document}
\title{Large Language Model Driven Agents for Simulating Echo Chamber Formation}
\titlerunning{Language Model Driven Agents for Simulating Echo Chamber Formation} 
\author{Chenhao Gu \and
Ling Luo \and
Zainab Razia Zaidi \and
Shanika Karunasekera
}

\authorrunning{C. Gu, L. Luo, Z. R. Zaidi, S. Karunasekera}

\institute{The University of Melbourne, Melbourne, Australia}

%
%
%
%
\maketitle              
\begin{abstract}
The rise of echo chambers on social media platforms has heightened concerns about polarization and the reinforcement of existing beliefs. Traditional approaches for simulating echo chamber formation have often relied on predefined rules and numerical simulations, which, while insightful, may lack the nuance needed to capture complex, real-world interactions. In this paper, we present a novel framework that leverages large language models (LLMs) as generative agents to simulate echo chamber dynamics within social networks. The novelty of our approach is that it incorporates both opinion updates and network rewiring behaviors driven by LLMs, allowing for a context-aware and semantically rich simulation of social interactions. Additionally, we utilize real-world Twitter (now X) data to benchmark the LLM-based simulation against actual social media behaviors, providing insights into the accuracy and realism of the generated opinion trends. Our results demonstrate the efficacy of LLMs in modeling echo chamber formation, capturing both structural and semantic dimensions of opinion clustering. 

\end{abstract}

\section{Introduction}

In recent years, the rise of echo chambers on social media platforms has become a significant concern, particularly due to their impact on polarization, misinformation, and the reinforcement of existing beliefs. Echo chambers, where individuals are exposed predominantly to information that confirms their preexisting opinions, can lead to social fragmentation and hinder healthy public discourse \cite{bonabeau2009decisions}. Understanding and simulating the dynamics of echo chambers is crucial for advancing our knowledge of how opinions evolve and cluster within online communities. Traditional models for studying echo chamber formation have largely relied on pure numerical simulations, using predefined behavioral rules to predict the evolution and clustering of opinions \cite{anderson2019recent}. While these models have contributed valuable insights, they often fall short of capturing the rich, complex interactions that occur in real-world social networks.

The rapid advancements in Large Language Models (LLMs) offer a new avenue for exploring these complex dynamics. LLMs, with their deep understanding of language and context, provide the potential to simulate not only the structure of social interactions but also the nuanced content and sentiment that drive echo chamber formation. In this paper, we propose a novel framework that leverages the interpretive power of LLMs to simulate the emergence of echo chambers. By integrating real data with LLM-driven simulations, our framework provides a more sophisticated and context-aware analysis of how opinions are shaped and reinforced within online communities. Our \emph{contributions} can be summarized as:

\textbf{Modeling Echo Chamber Formation with LLMs:} We demonstrate the ability of LLMs form echo chambers to capture both \textbf{network structure} and \textbf{semantic relationships}, enabling realistic and context-aware simulations.

\textbf{Real-World Data Integration:} By incorporating real-world social network data, we perform comparative analyses of simulated and observed echo chambers, showcasing the practical applications of LLM-driven simulations.

\textbf{Comparing LLM Performance:} We evaluate and compare different LLMs, providing insights into their relative strengths and weaknesses.

Overall, our contributions demonstrate the transformative potential of LLMs in understanding and simulating echo chambers, using real-world data to drive these simulations, and LLM-driven analysis to advance the field.

\section{Related Work}

\textbf{Echo Chambers Modeling.} The study of opinion dynamics and echo chambers in social networks has a rich history, with foundational models such as the Friedkin-Johnson Dynamics Model \cite{friedkin1990social}, which simulates opinion evolution based on intrinsic beliefs and social influence, and the Bounded Confidence Model by Deffuant et al. \cite{deffuant2000mixing}, which restricts interactions to individuals within a confidence range. More recent works, such as Sasahara et al. \cite{sasahara2021social}, emphasize the dynamic interplay of opinion polarization and structural adjustments, such as unfollowing, and demonstrate how these behaviors accelerate the formation of echo chambers. Our work builds on these models by incorporating LLMs to simulate user behavior and opinion dynamics, leveraging advanced natural language understanding to provide deeper insights into the formation and evolution of echo chambers.

\textbf{LLM-Driven Social Simulation.} Leveraging LLMs for social simulation has emerged as a novel research direction, enabling both downstream task facilitation and a deeper understanding of LLM capabilities. Early work, such as Generative Agents \cite{park2023generative}, explored LLM-empowered agents that simulate human behavior in interactive environments. This idea was extended to social networks with systems like S3 \cite{gao2023s}, which simulate emotions, attitudes, and interactions to study population-level phenomena like information and emotion propagation. Furthermore, there have been attempts to use LLMs for simulating echo chambers \cite{wang2024decoding}, which differ from our approach in two key aspects: we argue that echo chambers partially stem from changes in network structures, and we emphasize that simulating echo chambers does not imply that more polarization is always better. Instead, we advocate for validating the performance of LLM-based simulations through comparisons with real-world data. 

\section{Problem Statement and Definitions}

This study aims to simulate the formation of echo chamber by capturing the evolution of user opinions and the structural changes in social networks. 
\vspace{-0.4cm}
\subsection{Opinion Evolution}
Each user \(i\) holds an opinion \(O_i(t)\) at time \(t\), influenced by their neighbor set \(N_i(t)\), representing their current social connections. The opinion update rule is inspired by the classical DeGroot model \cite{degroot1974reaching}, where opinions evolve as a weighted average of neighboring opinions. The general update rule is:
{\small
\[
O_i(t+1) = O_i(t) + \frac{1}{|N_i(t)|} \sum_{j \in N_i(t)} f(O_j(t), O_i(t)),
\]
}
where \(f(O_j, O_i(t))\) quantifies the impact of neighbor \(j\)'s opinion \(O_j\) on user \(i\). In the simplest form, \(f(O_j, O_i(t)) = w_{ij}(O_j - O_i(t))\), where \(w_{ij}\) is the weight of influence assigned to neighbor \(j\).
\vspace{-0.4cm}
\subsection{Network Rewiring}
Users dynamically adjust their connections over time based on opinion compatibility and recommendations, reflecting real-world follow/unfollow behaviors.

The probability of severing a connection (unfollow) or forming a new connection (follow) is determined by the compatibility function \(g\), defined as:
{\small
\[
P_{\text{unfollow}}(i, j) \propto 1 - g(O_i, O_j), \quad P_{\text{follow}}(i, k) \propto g(O_i, O_k),
\]
}
where \(0 \leq g(O_i, O_j) \leq 1\) quantifies the similarity between user \(i\)'s and \(j\)'s opinions. The neighbor set \(N_i(t)\), representing user \(i\)'s incoming connections in the directed graph, evolves as these follow/unfollow actions modify the graph.
\vspace{-0.4cm}
\subsection{LLM-enhanced Approach}
\label{sec:llm_approach}
In the LLM-enhanced framework, the influence function \(f(O_j, O_i(t))\) and the compatibility function \(g(O_i, O_j)\) are defined through prompt-driven models, incorporating textual context into the simulation. Specifically, the functions are reformulated as:
{\small
\[
f(O_j, O_i(t)) = \text{LLM}(T_o(C_i, C_j)), \quad g(O_i, O_j) = \text{LLM}(T_r(C_i, C_j)),
\]
}where \(T_o\) (Prompt template for opinion dynamic) and \(T_r\) (Prompt template for rewiring compatibility) serve as input structures to instruct the LLM on specific tasks. The input \(C_i\) represents the historical content of user \(i\), while \(C_j\) represents the historical content of user \(j\). Here, ``historical content'' typically refers to a subset of the user's recent posts, such as their 10 most recent posts. 

Additionally, the LLM generates a new post \(y_i(t)\) for user \(i\), reflecting their opinion and the surrounding context. This process is defined as:
{\small
\[
y_i(t) \sim \text{LLM}(T_g(C_i, \{C_j \mid j \in N_i(t)\})),
\]
}where \(T_g\) (template for content generation) combines \(C_i\) with the contexts \(\{C_j\}\) from neighbors in \(N_i(t)\) to produce context-aware content. These template structures ensure that the LLM can dynamically adapt to the textual and contextual information, unifying opinion dynamics, rewiring decisions, and content generation into a cohesive framework.




\section{Model Framework}
\begin{figure}
    \centering
    \includegraphics[width=\textwidth]{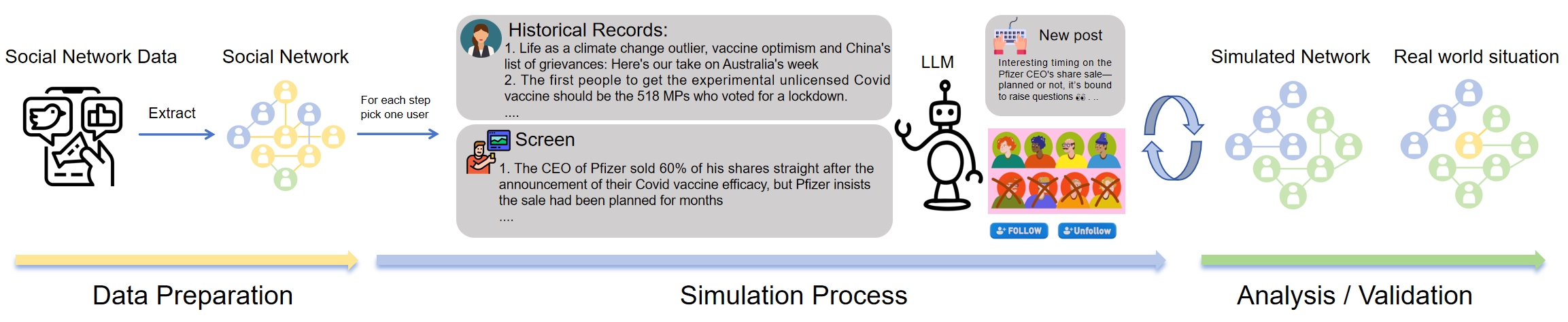}
    \caption{\small The framework is divided into three main components: (1) \textbf{Data Preparation}, where social network data is collected and used to build an user network; (2) \textbf{Simulation Process}, where an LLM generates user posts and adjusts connection dynamically; and (3) \textbf{Analysis and Validation}, which analyzes the simulated structure of user interactions and echo chamber effect.}
    \label{framework}
\end{figure}

The objective of this study is to develop a simulation framework that models echo chamber dynamics with theoretical and practical alignment to real-world social network behaviors, including opinion distribution and network structure dynamics. The framework captures the interactions and structural changes in social networks, encompassing opinion evolution, network rewiring, and content generation, as outlined in Figure~\ref{framework}.

\vspace{-0.8cm}
\subsection{Data Preparation}
The framework primarily uses Twitter data due to its rich interaction patterns, though it supports datasets from other platforms as well. Key inputs include user interactions, posts, and follow relationships, which are used to construct an active user network.

To enhance efficiency and relevance, the framework prioritizes highly active users, motivated by two key considerations: (1) the computational infeasibility of simulating the entire network due to the substantial resources required, and (2) the limited interactions of infrequent users, which are insufficient for meaningful simulation and may introduce noise into the analysis. By prioritizing active users, the simulation captures meaningful network dynamics.

The constructed network comprises nodes (users) with distinct opinions and edges (social connections) based on social interactions, such as retweets on Twitter. This network evolves over time to reflect the dynamic of social interactions.
\vspace{-0.3cm}
\subsection{Simulation Process}

The simulation captures the dynamic user opinions and network connections in social networks. At each iteration, a randomly selected user refreshes their feed, viewing a limited subset of prioritized posts from connected ``friends'' or recommended content. This process, referred to as the \textbf{screen}, represents the information accessible to users via social media and simulates their finite cognitive capacity in real-world interactions.

Based on the screen and the user's historical activity, the LLM generates a new post that reflects the user's updated opinion. Subsequently, the user evaluates their network connections, probabilistically disconnecting from those with conflicting opinions and forming new connections with like-minded individuals. This adaptive process, as outlined in Section~\ref{sec:llm_approach}, is driven by the LLM and mirrors real-world social interactions where network structures evolve based on opinion alignment.



Over time, the network exhibits clustering of users with similar opinions, while those with conflicting opinions become isolated. The simulation concludes when the network reaches a stable state, characterized by stabilized opinions and connections, or after a predefined number of iterations. The simulation Process component of Figure \ref{framework} illustrates how historical data and network information contribute to opinion updates and connection adjustments.

\subsection{Analysis and Validation}

The final component of the framework focuses on validation and analysis to ensure the simulation's relevance and reliability. This process involves a qualitative comparison between the simulated network and real-world social network data across various metrics. The goal is to evaluate how well the LLM-driven simulation aligns with real-world social network phenomena, identifying both consistencies and discrepancies.

This approach is conceptually similar to a \emph{digital twin}, where the simulation serves as a virtual counterpart to the initial real-world social network. By analyzing how the simulated network evolves under LLM influence, researchers can gain insights into behavioral dynamics, such as opinion formation, clustering, and network rewiring, and compare them with real-world trends.


\section{Prompt Templates}

To ensure reproducibility and enable systematic exploration of different prompt designs, the framework employs standardized prompting mechanisms. These prompts guide the LLM to model user behaviors effectively while maintaining consistency across tasks. The prompt design adheres to two key principles: \textbf{Standardization.} Prompts follow a consistent format with explicitly defined tasks and output specifications. This ensures clarity and reproducibility across simulations. \textbf{Chain-of-Thought Reasoning \cite{wei2022chain}.} Prompts encourage step-by-step reasoning to enhance the interpretability and reliability of the generated outputs.

\begin{figure}
    \centering
    \includegraphics[width=0.7\textwidth]{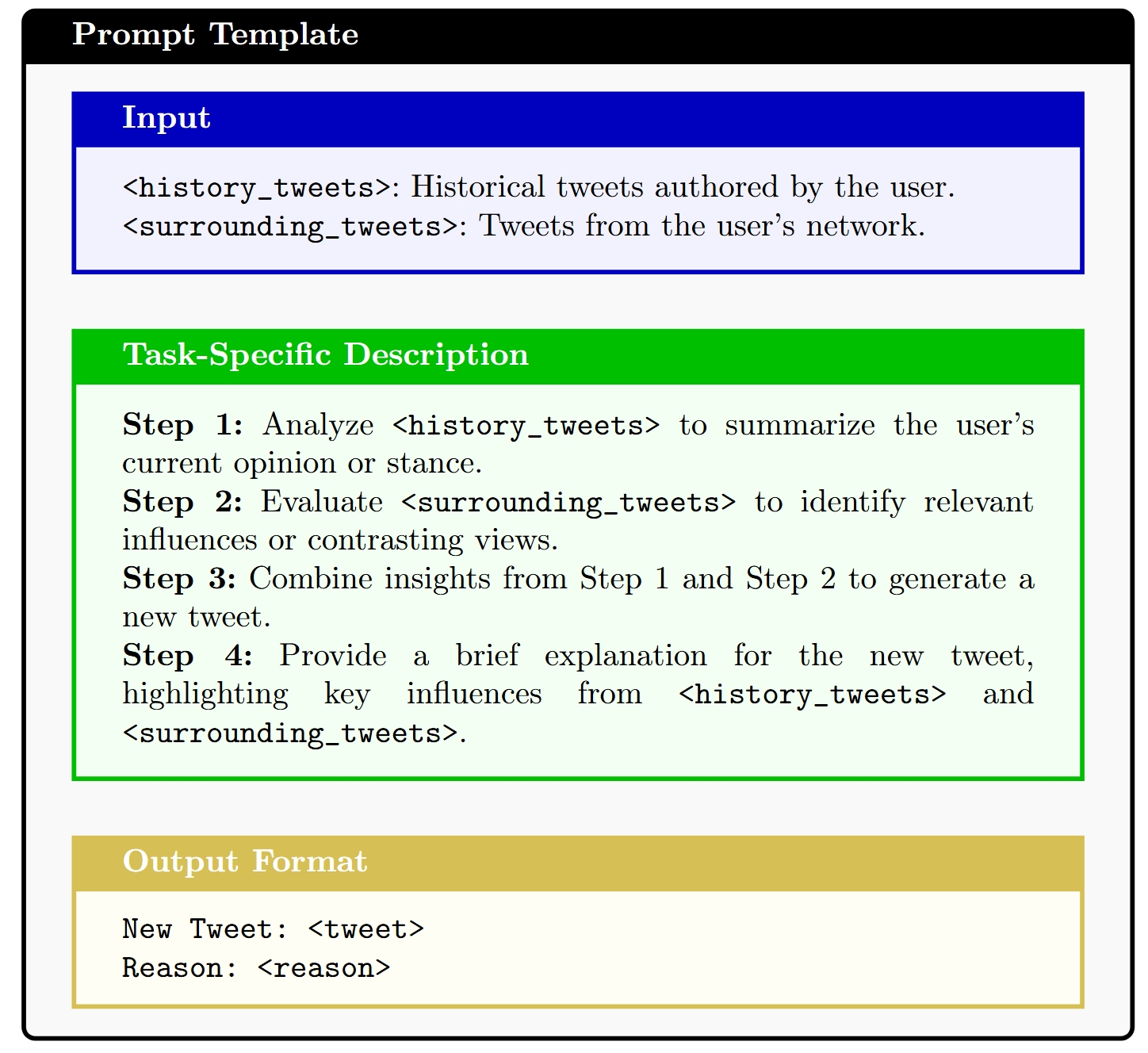}
    \caption{\small Prompt template example}
    \label{template}
\end{figure}

An example prompt template for generating a new post based on a user’s history and surrounding tweets is shown in Figure \ref{template}. This template can be adapted for other tasks, such as opinion updates or network rewiring, by modifying the task-specific instructions and input contexts. The consistent use of structured prompts facilitates future exploration of prompt designs and their effects on simulation outcomes.

\vspace{-0.5cm}
\section{Experiments}
\subsection{Stand Alone Simulation}

In this section, we present a pure simulation experiment using ChatGPT and Gemini to explore the formation of echo chambers. This feasibility test aims to demonstrate how LLMs simulate the evolution and divergence of opinions, ultimately leading to distinct clusters of like-minded users.
\begin{figure}
    \centering
    \includegraphics[width=0.5\textwidth]{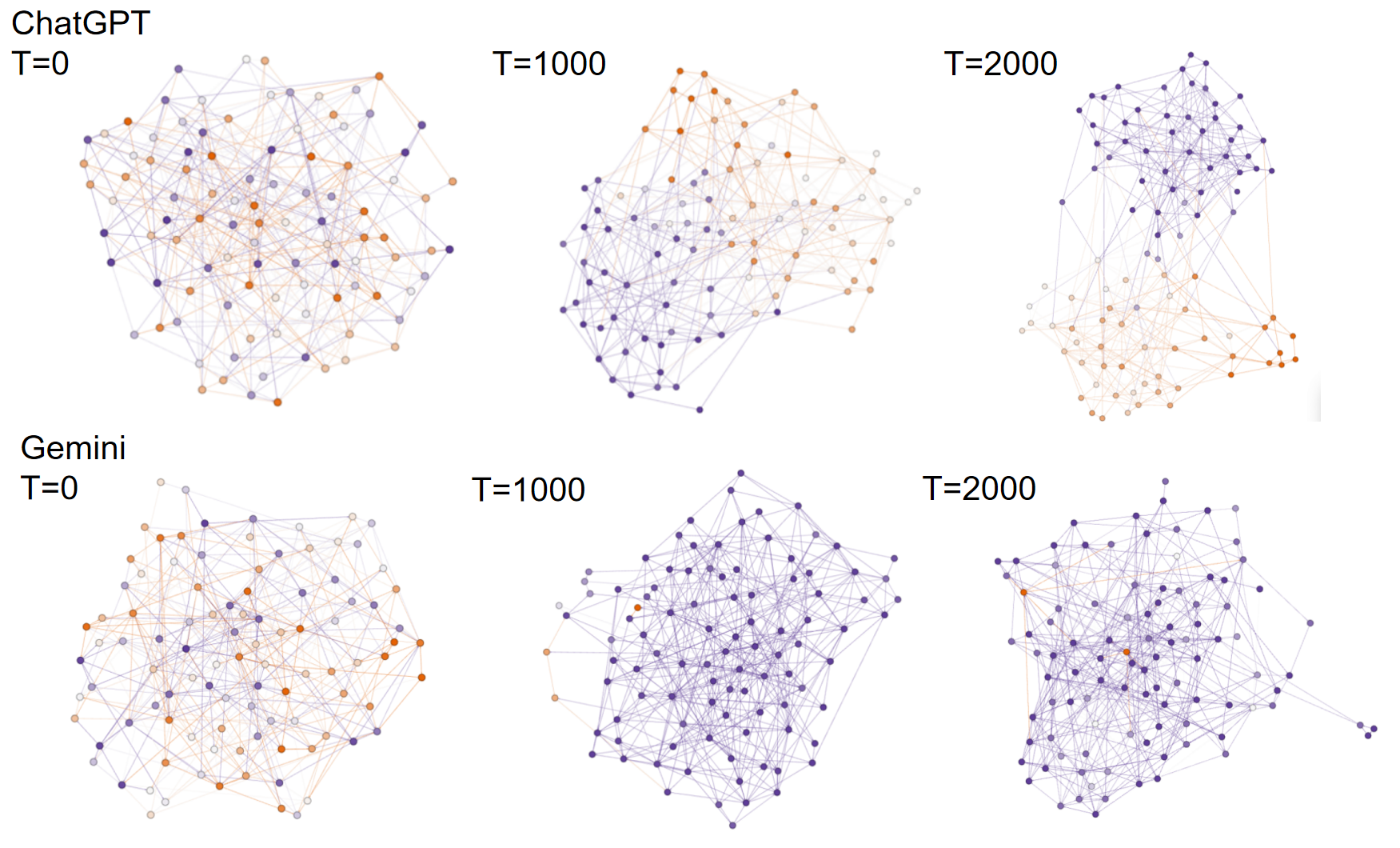}
    \caption{\small Simulation results using ChatGPT and Gemini. Opinion dynamics at different time steps illustrate the gradual formation of echo chambers.}
    \label{purly}
\end{figure}

Figure~\ref{purly} visualizes the simulation, where node colors represent opinion states: shades transition from orange (-1) to deep blue (1), with intensity reflecting the degree of agreement or disagreement. Initially, opinions are randomly distributed across the network. Over time, the influence of neighboring nodes causes opinions to cluster, changing the network structure into distinct echo chambers.

Notably, in the simulations, ChatGPT tends to produce highly polarized results, with opinions splitting into opposing clusters. In contrast, Gemini exhibits a stronger tendency toward consensus, often resulting in more uniform positive sentiment. These observations suggest potential biases in LLMs, likely influenced by differences in their training methods and datasets. Given that our topic focuses on COVID-19 vaccines, it is possible that Gemini has been fine-tuned to generate more positive attitudes toward the subject. Additionally, the starting state of the simulation significantly impacts the final outcome. Since these simulations operate in a purely hypothetical scenario, we lack a definitive standard to evaluate which simulation is more accurate or realistic. This underscores the importance of integrating real-world data to validate and enhance the credibility of LLM-driven simulations.


\subsection{Real data-driven simulation}

In this section, we outline the experimental setup designed to evaluate our proposed framework for modeling echo chamber formation using real-world data. We provide a detailed description of the datasets, simulation parameters and evaluation metrics employed to analyze opinion dynamics and echo chamber formation across different contexts.

The experiments are conducted on two datasets:
\begin{enumerate}
    \item \textbf{COVID-19 Vaccination Dataset:} This dataset was collected from global English-language tweets using COVID-19-related keywords . Details about the keywords and the collection process can be found in \cite{Lamsal2021,zaidi2023topics}. The dataset includes tweets with stances categorized as \textit{favoring} vaccination (pro-vax), \textit{opposing} vaccination (anti-vax), or \textit{neutral}, focusing on the topic of vaccine hesitancy \cite{zaidi2023topics}.
    \item \textbf{Ukraine War Dataset:} Introduced in \cite{perera2024quantifying}, this dataset includes 1 million English-language tweets collected using a keyword search between August 1 and August 31, 2022. Each tweet's stance was labeled as \textit{pro-Russian}, \textit{pro-Ukrainian}, or \textit{neutral}, providing a rich resource for studying polarized discourse.
\end{enumerate}


\subsubsection{Model Selection and Experimental Setup}

To evaluate the simulation of opinion dynamics, we utilized six generative models and one traditional numerical simulation, facilitating a comparison between LLM-driven and rule-based approaches.

The LLM-based models included both closed-source and open-source options, offering diverse capabilities. Closed-source models include \textbf{ChatGPT} (OpenAI), \textbf{GPT4o Mini} (OpenAI), and \textbf{Gemini} (DeepMind), while open-source models include \textbf{Gemma} (DeepMind) Gemma2-27b version, \textbf{Meta-Llama} (Meta) Llama3.1 70B Instruct version, and \textbf{Qwen} (Alibaba) Qwen2 72B Instruct version. These models varied in design, with strengths ranging from contextual reasoning to domain adaptability. As a baseline, we used a traditional \textbf{Equation-based simulation} \cite{sasahara2021social}, relying on predefined mathematical rules to model opinion interactions. This served as a control to benchmark the added value of LLMs in capturing semantic and contextual nuances.

Given the inherent randomness of both the models themselves and the outputs generated by LLMs, each simulation was run five times per model, and the results were averaged. The experiments were initialized using 6000 tweets records, with the network state evaluated after generating 2,000 simulated steps.

\subsubsection{Evaluation Metrics}

We assessed model performance using the following metrics:  
\textbf{Modularity \cite{newman2004finding}:} Evaluates community structures, where high values indicate distinct clusters characteristic of echo chambers.  
\textbf{Clustering Coefficient \cite{botte2022clustering}:} Measures local connectivity, reflecting the degree of tightly interconnected groups.  
\textbf{Opinion Accuracy:} Measures model predictions to real-world stances, gauging alignment with observed data.  
\textbf{Average Path Length and Network Density:} Analyze the overall network structure, including user connectivity and information propagation.

These metrics provide a comprehensive evaluation framework for assessing how well LLM-based simulations capture opinion polarization and echo chamber dynamics.

\section{Experimental Result Analysis}
\subsection{Network Metrics}

\begin{table}[htbp]
\centering
\scriptsize 
\caption{\small Comparison of LLMs Based on Network Metrics. The arrows (\(\uparrow\), \(\downarrow\)) indicate the direction of change relative to the real data. Bold values highlight the models whose results are closest to the real data.}
\label{tab:llm_network_metrics} 
\resizebox{\textwidth}{!}{%
\begin{tabular}{clllll}
\toprule
\textbf{Dataset} & \textbf{Model} & \textbf{Modularity} & \textbf{Clustering} & \textbf{Path Length} & \textbf{Density} \\
\midrule
\multirow{8}{*}{\centering COVID-19} 
& Real Data & 0.7125 & 0.0393 & 5.1384 & 0.0024 \\
\cmidrule{2-6}
& GPT4o Mini & 0.7055 $\downarrow$ & 0.0461 $\uparrow$ & 4.9221 $\downarrow$ & 0.0015 $\downarrow$ \\
& ChatGPT & 0.6855 $\downarrow$ & 0.0739 $\uparrow$ & 4.8250 $\downarrow$ & \textbf{0.0016} $\downarrow$ \\
& Gemini & 0.7189 $\uparrow$ & 0.0585 $\uparrow$ & 4.9103 $\downarrow$ & 0.0015 $\downarrow$ \\
& Meta-Llama & 0.7257 $\uparrow$ & 0.0542 $\uparrow$ & 5.0167 $\downarrow$ & 0.0015 $\downarrow$ \\
& Qwen & \textbf{0.7096} $\downarrow$ & 0.0631 $\uparrow$ & 5.0718 $\downarrow$ & \textbf{0.0016} $\downarrow$ \\
& Gemma & 0.7299 $\uparrow$ & \textbf{0.0431} $\uparrow$ & \textbf{5.1850} $\uparrow$ & \textbf{0.0016} $\downarrow$ \\
& Equation & 0.6878 $\downarrow$ & 0.0267 $\downarrow$ & 5.5711 $\uparrow$ & \textbf{0.0016} $\downarrow$ \\
\midrule
\multirow{8}{*}{\centering Ukraine War} 
& Real Data & 0.3706 & 0.0853 & 2.9399 & 0.0026 \\
\cmidrule{2-6}
& GPT4o Mini & 0.3764 $\uparrow$ & 0.1160 $\uparrow$ & 2.8146 $\downarrow$ & \textbf{0.0026}  \\
& ChatGPT & \textbf{0.3660} $\downarrow$ & 0.0690 $\downarrow$ & \textbf{2.8550} $\downarrow$ & \textbf{0.0026} \\
& Gemini & 0.3837 $\uparrow$ & 0.1162 $\uparrow$ & 2.8269 $\downarrow$ & \textbf{0.0026}  \\
& Meta-Llama & 0.3792 $\uparrow$ & 0.1138 $\uparrow$ & 2.8415 $\downarrow$ & \textbf{0.0026}\\
& Qwen & 0.3800 $\uparrow$ & 0.1219 $\uparrow$ & 2.8185 $\downarrow$ & \textbf{0.0026}  \\
& Gemma & 0.3802 $\uparrow$ & 0.1151 $\uparrow$ & 2.8251 $\downarrow$ & \textbf{0.0026}  \\
& Equation & 0.4137 $\uparrow$ & \textbf{0.0717} $\downarrow$ & 3.1932 $\uparrow$ & \textbf{0.0026} \\
\bottomrule
\end{tabular}%
}
\end{table}

Table \ref{tab:llm_network_metrics} compares network metrics for different LLMs and traditional methods on the COVID-19 and Ukraine War datasets. We evaluate the models' ability to replicate key structural properties related to echo chamber formation using modularity, clustering coefficient, average path length, and network density. The COVID-19 dataset reflects entrenched polarization, while the Ukraine War dataset shows less division due to a dominant pro-Ukraine community. It is important to note that in these simulations, a stronger echo chamber effect is not always desirable. Our aim is for the simulated results to closely match the patterns observed in real data. For instance, some models may exhibit very high modularity, which could be an extreme outcome and not necessarily beneficial, depending on the purpose of the simulation.

For the COVID-19 dataset, real-world data shows high modularity (0.7125) and clustering structure, indicative of echo chambers in vaccine discourse. In LLM simulations, certain models like Meta-Llama (0.7257) and Gemma (0.7299) further amplify modularity, suggesting exaggerated division, while ChatGPT (0.6855) exhibits weaker polarization, reflecting a less distinct community structure compared to real-world data. Clustering coefficients increase in LLM simulations, highlighting strong local cohesion, but average path lengths decrease (e.g., ChatGPT: 4.8250 vs. real data: 5.1384), suggesting faster information spread. Network density uniformly decreases across different models, reflecting sparser connections in simulations.

In the Ukraine War dataset, real-world data shows lower modularity (0.3706) due to a dominant pro-Ukraine community. LLM-based models, such as Gemini (0.3837) and Meta-Llama (0.3792), enhance modularity and clustering, creating more distinct communities. However, traditional methods like Equation (0.4137) tend to overemphasize the formation of echo chambers, resulting in more polarized communities that deviate significantly from the lower modularity observed in real-world data. Average path lengths decrease across models (e.g., Gemini: 2.8269 vs. real data: 2.9399), indicating efficient information diffusion. Network density remains consistent (0.0026), preserving overall connectivity.

\textbf{Discussion} The findings highlight the clear advantages of LLM-based simulations over traditional methods like Equation-based simulation in modeling complex social networks. LLMs demonstrate superior adaptability, effectively capturing the distinct structural properties of datasets with varying initial conditions. For the COVID-19 dataset, they replicate the entrenched modularity and clustering of real-world networks, while for the Ukraine War dataset, they enhance community distinction and local connectivity, even when starting from a less structured network.

One of the key takeaways is the ability of LLMs to balance modularity and clustering. While higher modularity indicates more distinct community formation, excessively high values, such as those produced by Meta-Llama and Gemini, may reflect overfitting to echo chamber dynamics. Conversely, models like ChatGPT offer a more balanced approach, aligning more closely with real-world data while maintaining meaningful structural insights. This balance is critical for simulations that aim to realistically represent social network dynamics rather than amplify existing biases.

The consistent underperformance of Equation underscores the limitations of traditional methods in capturing the complexity of real-world social networks. The method's inability to achieve realistic clustering and modularity, particularly for the Ukraine War dataset, suggests that it lacks the nuanced understanding required to model evolving network structures. In contrast, LLMs leverage their advanced language understanding and contextual capabilities to simulate both entrenched and emergent echo chambers effectively.

\subsection{Track Stance Dynamics}

In this section, we evaluate the LLM simulators' performance in tracking the stance dynamics in real-world data. By using real data as a benchmark, we can compare the accuracy of different simulation models in replicating observed opinion trends.

\begin{figure}
    \centering
    \includegraphics[width=0.7\textwidth]{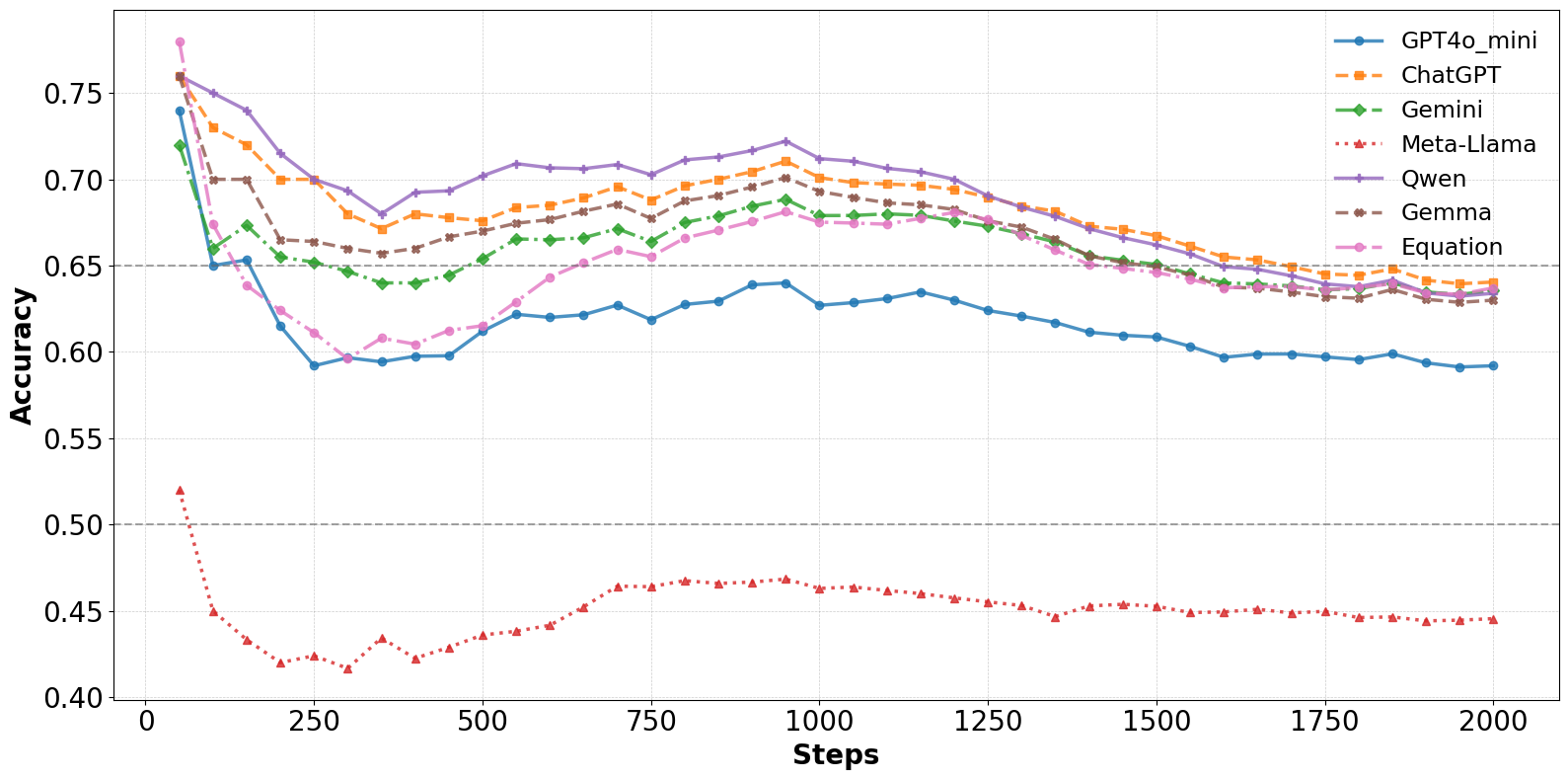}
    \caption{\small Comparison of stance accuracy across different models in COVID-19 dataset} 
    \label{accuracy_analysis}
\end{figure}

The evaluation revealed notable variability in the performance of LLMs. Among the models, ChatGPT and Qwen consistently demonstrated high accuracy, effectively capturing nuanced stance dynamics while maintaining stability across training steps. In contrast, Meta-Llama exhibited significantly lower performance, primarily due to inconsistent output formatting. For example, unexpected line breaks or irregular word segmentation often occurred, which disrupted the accurate extraction of stance values. As the same prompt was employed for all models, these formatting issues highlight intrinsic limitations of the Meta-Llama model, which need further attention.

The equation-based method exhibited distinct performance characteristics, with low accuracy and high fluctuations in the early stages before eventually stabilizing at a moderate level. This trajectory underscores the limitations of simpler models in adapting to the complexities of dynamic stance evolution without the benefit of contextual embeddings. Nevertheless, the eventual stabilization suggests some alignment with observed data trends over time.

\subsection{Language Distribution Analysis}

Using LLMs to generate natural language tweets represents a significant advantage in LLM-based simulations, marking a breakthrough compared to previous simulations that lacked this generative capability. To evaluate the quality of the generated language, we compared simulated and real-world tweets by transforming both datasets into embeddings using the sentence transformer model \cite{reimers2019} and performing clustering to analyze language distribution and structural differences.
\begin{figure}[h]
    \centering
    \includegraphics[width=0.7\textwidth]{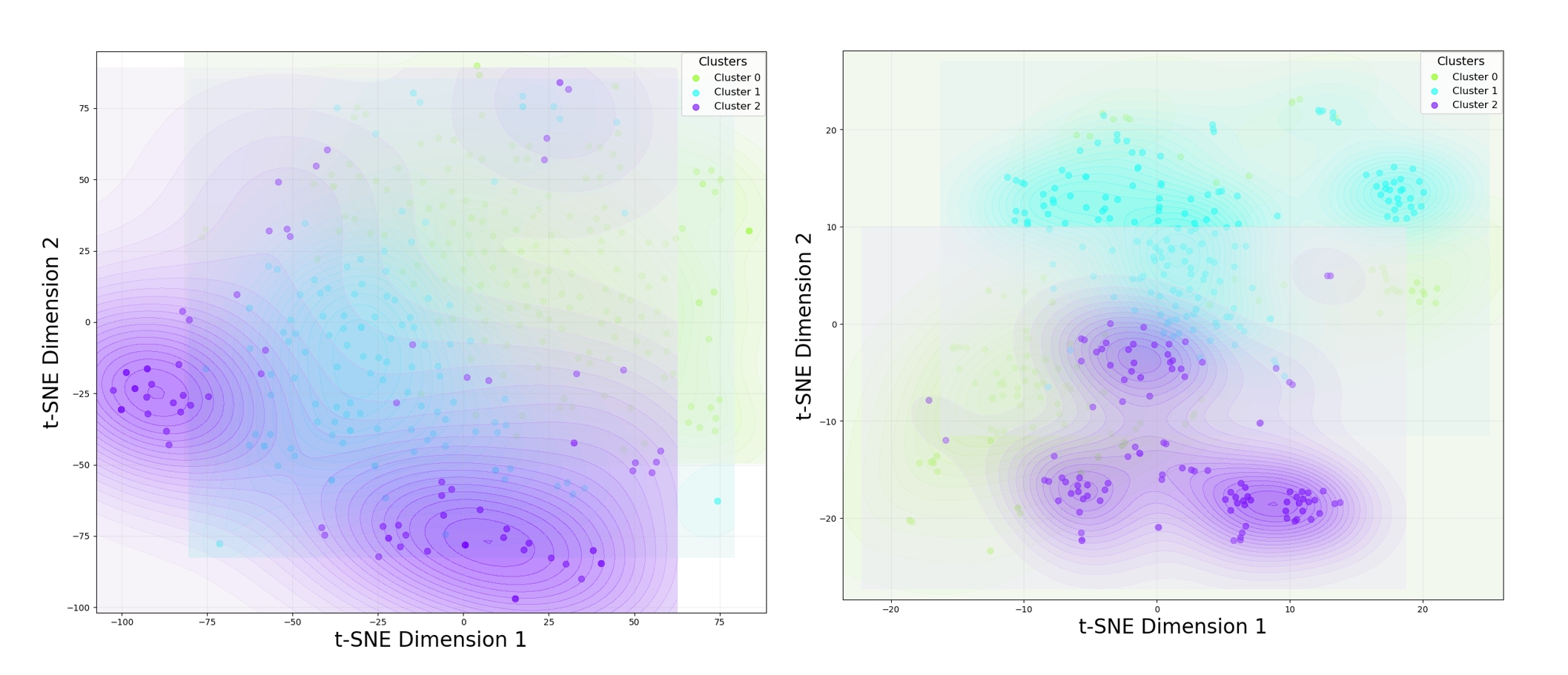}
    \caption{\small Clustering of real-world data (left) and simulated data (right) illustrated in t-SNE embedding space.}
    \label{fig:language_entropy}
\end{figure}

Figure~\ref{fig:language_entropy} presents the clustering results for real-world and simulated tweet embeddings. Simulated tweets demonstrate more concentrated semantics, forming tighter clusters with a silhouette score of 0.0966 compared to 0.0485 for real-world tweets and an average intra-cluster distance of 0.5321 versus 0.7456. While this indicates that simulations capture cohesive trends, they also reveal less linguistic diversity and dispersion than real-world tweets, as shown by the smaller average inter-cluster distance (0.4354 vs. 0.3712). This suggests that although simulated tweets exhibit certain clustering structures, they lack the broader variability and richness found in real-world data.
\section{Conclusion}

In this study, we explored the use of LLMs as generative agents for simulating echo chamber formation in social networks. Leveraging the interpretive power of LLMs, our framework not only captures complex social dynamics, opinion interactions, and realistic user-generated content but also tracks the evolving structure of network changes, offering a significant improvement over traditional numerical simulations. Experiments demonstrated that LLMs provide a closer approximation of real-world echo chambers, effectively simulating nuanced opinion updates, rewiring decisions, and stance predictions. By using real-world data as a validation benchmark, our model ensures greater alignment with actual social media dynamics and highlights its robustness in replicating real-world patterns. In future work, we can incorporate diverse training data or refined prompt engineering, to more effectively replicate the linguistic complexity of real-world social media interactions.

\bibliographystyle{splncs04} 
\bibliography{references} 

\end{document}